\documentclass[showpacs,twocolumn,preprintnumbers]{revtex4-1}
\usepackage{indentfirst}
\usepackage{graphicx}
\usepackage{amsmath}
\usepackage{amsfonts}
\usepackage{color}
\usepackage[sc]{mathpazo}
\begin{document}
\preprint{Published: Phys. Lett. A 376 (2012) pp. 1791-1796}

\title{Creation of quantum correlations between two atoms in a dissipative environment from an initial vacuum state}
\author{Ferdi Altintas}\email{altintas\textunderscore f@ibu.edu.tr} \author{Resul Eryigit}\email{resul@ibu.edu.tr}
\affiliation{Department of Physics, Abant Izzet Baysal University, Bolu, 14280, Turkey.}
\begin{abstract}
We have investigated the effect of counter-rotating terms on the dynamics of entanglement and quantum discord between two identical atoms interacting with a lossy single mode cavity field for a system initially in a vacuum state. The counter-rotating terms are found to lead to steady states in the long time limit which can have high quantum discord, but have no entanglement. The effect of cavity decay rate on this steady state quantum discord has been also investigated, surprisingly, the increase in cavity decay rate is found to both enhance  and maximize the steady quantum discord for separable states.
\end{abstract}

\maketitle
\section{Introduction}
Rabi model is one of the most complete quantum mechanical models used to describe the interaction between a bosonic field and a two-level system~\cite{iirabi}, such as a two-level atomic system interacting with an electromagnetic field in a cavity in quantum optics~\cite{aljhe}, an electronic spin coupled to phonon modes of a crystal lattice~\cite{holstein},  superconducting qubits interacting with a nanomechanical resonator~\cite{scqnmr}, and many more examples can be found in Ref.~\cite{etcref}. The Rabi Hamiltonian reads~($\hbar=1$)
\begin{eqnarray}\label{rabiham}
H=\frac{\omega_0}{2}\sigma_z+\omega a^{\dagger}a+g(\sigma_++\sigma_-)(a+a^{\dagger}),
\end{eqnarray}
where $\omega$ and $\omega_0$ are the field and atomic transition frequencies, respectively, $g$ is the coupling constant, $a$~$(a^{\dagger})$ is the annihilation~(creation) operator of the bosonic field and $\sigma_z$, $\sigma_+$ and $\sigma_-$ are the pseudo-spin operators. It has been difficult to find an analytic solution to the Rabi model. Recently, Braak has shown that its eigenvalues can be calculated as the roots of a polynomial~\cite{braak}. However, a large number of studies on the model involve numerical~\cite{nrm}, perturbative~\cite{prm} and approximate analytical solutions~\cite{arm}. The most widely used approximation is the so called rotating wave approximation~(RWA) which amounts to ignoring the counter-rotating terms,~$a^{\dagger}\sigma_+$ and $a\sigma_-$, in the field-atom interaction~\cite{arm}. The ignored terms correspond to the emission and absorption of virtual photons without energy conservation. The error introduced by RWA depends on the magnitudes of the atom-field frequency detuning and the atom-field interaction strength. This approximation is thought to be valid only for small atom-field frequency detuning~(i.e., $|\Delta|=|\omega_0-\omega|<<\omega,\omega_0$) and weak couplings~(i.e., $g<<\omega,\omega_0$). In general, RWA is justified in typical optical setups,  because the atom-field coupling strength divided by the field frequency is of the order  $10^{-7}-10^{-6}$ and the nearly resonant condition can be satisfied~\cite{hldpk}.  In this limit, Rabi model is known as Jaynes-Cummings model which can be integrated exactly, so it is attractive and has been used successfully for more than four decades to explain many physical phenomena in quantum optics, such as Rabi oscillations~\cite{burne}, squeezing~\cite{jrkjlm}, non-classical states~(such as Schr\"{o}dinger cat-like states)~\cite{burne2}, Fock states~\cite{wvhw}, collapse and revival of atomic inversion~\cite{grhw} and quantum as well as classical correlations between atom-atom or atom-field systems~\cite{gglxg}. However, recent developments in physical implementation of qubits lead to systems with $g/\omega$ up to one and possibly much higher than one which requires a careful consideration of the effect of virtual processes~(counter-rotating terms) on atom-field interaction~\cite{aljhe,holstein,scqnmr,etcref}.

The quantum correlations are crucial to many quantum information tasks; for example entanglement is important in speeding up quantum algorithms, making quantum teleportation~\cite{bw} and cryptography~\cite{ekert} possible. On the other hand, quantum discord~(QD)~\cite{howz} has a role in the deterministic quantum computation with one pure qubit~\cite{dqc1} which was also demonstrated experimentally in recent times~\cite{exdqc1} and in Grover search algorithm~\cite{grover}. Dakic {\it et al.} recently showed that QD is an optimal resource for remote quantum state preparation~\cite{optimal}. However, the role of QD in speeding up quantum operations is still controversial. On the other hand, all realistic quantum systems are inevitably connected to decohering and/or dissipative environments which wipes out the quantumness of the systems. The effect of decoherence and dissipation on the dynamics of quantum discord and entanglement between two qubits in contact with an environment have been investigated very throughly by many groups in the last decade experimentally as well as theoretically~\cite{gglxg,tyjhe,esdexp,farejpa,fwbac,wsfb,fare1,aucca,fare2,mcsv,mwfcs,mps,hxyz,fcvg}. It was shown that the entanglement is very fragile and can cease to exist in a finite time, although the coherence of the single qubit decays exponentially. This effect is named as entanglement sudden death~(ESD)~\cite{tyjhe} and also observed experimentally~\cite{esdexp}. In contrast to entanglement, QD is more robust under decoherence and dissipation and it is considered as a measure of more-general-than-entanglement type quantum correlations. Actually, it was shown by Ferraro {\it et al.} that almost all quantum states of a bipartite system have non-zero quantum discord~\cite{ferraro}. It was found that QD presents an instantaneous disappearance at some time points in non-Markovian regime~\cite{fwbac} and asymptotic decay in Markovian regime even at finite temperatures~\cite{fwbac,wsfb}, while entanglement dies in a finite time. In Refs.~\cite{mps,hxyz}, QD is found to be unaffected by decoherence or dissipation for certain initial states contrary to the entanglement dynamics which suffers sudden death. In Refs.~\cite{fare1,fcvg}, quantum correlations, as quantified by quantum discord,  are found to increase under memoryless dissipation for some initially separable states, although the qubits remain unentangled for all time.

The counter-rotating terms are found to be responsible for several novel quantum mechanical effects, for example generation of photons as well as entanglement from states with no initial excitation~(vacuum) is such subject which has received plenty of attention in recent years~\cite{wddv,fjl,nb,swkhk,dodonov}. Along these lines,  Ficek {\it et al.} showed that it is possible to create high degree of entanglement between two qubits in a leaky cavity with the help of the virtual processes~\cite{fjl}.  An efficient method to generate entanglement between two separate ensembles of molecules trapped inside a superconducting resonator through which they are strongly coupled to a microwave field mode has been outlined in Ref.~\cite{nb}. The possibility of creation of photons as well as atom-cavity entangled states in the experimental implementation of the Landau-Zener sweeps~\cite{swkhk} and the non-stationary circuit QED~\cite{dodonov} was proposed recently. The creation of more-general-than-entanglement type quantum correlations, such as quantum discord, from initially vacuum states is an interesting and significant topic and  has not received enough attention. To the best of our knowledge, there is no reported study concerning the creation of QD from initially vacuum states.

In this Letter, we have studied the creation of entanglement and quantum discord between two qubits interacting with a lossy single mode cavity field and initially in a state with zero excitation. We have demonstrated that the counter-rotating terms in the Rabi Hamiltonian lead to steady states in the long-time limit  which have no or negligibly small entanglement but can have high quantum discord. It is natural to expect that the cavity decay decreases the magnitude of the steady state quantum discord, but the results we have obtained indicate that an increase in cavity decay rate increases and also maximizes the magnitude of the steady state quantum discord for separable states when the virtual processes are taken into account.
\section{The Model}\label{model}
Here, we consider two noninteracting two-level atoms, called $A$ and $B$, coupled to a single mode cavity field in a leaky cavity. The dynamics of the atoms and the cavity field is determined by the master equation for the density operator $\rho_S$ of the atoms plus cavity field~(for $\hbar=1$)~\cite{wddv,fjl,jlf}:
\begin{eqnarray}\label{mastereqn}
\dot{\rho}_S=-i[H,\rho_S]-\frac{\kappa}{2}(a^{\dagger}a\rho_S-2a\rho_S a^{\dagger}+\rho_S a^{\dagger}a),
\end{eqnarray}
where $\kappa$ is the damping rate of the cavity mode and determines the quality of the cavity with the relation $Q=\omega/\kappa$, where $\omega$ is the field frequency. $H$ in Eq.~(\ref{mastereqn})  is the Hamiltonian of the atom-cavity system including counter-rotating terms and can be obtained by replacing $\sigma_z$, $\sigma_+$ and $\sigma_-$ with $(\sigma_z^A+\sigma_z^B)$, $(\sigma_+^A+\sigma_+^B)$ and $(\sigma_-^A+\sigma_-^B)$ in Eq.~(\ref{rabiham}), respectively. Here, we have assumed that the atoms have identical transition frequencies, $\omega_0$, and coupled to the field mode with the same coupling constant, $g$.

The interaction terms in the Hamiltonian for two atoms interacting with a single-mode cavity field contain terms such as $a^{\dagger}\sigma_+^j$ and $a\sigma_-^j$ (for $j=A,B$) which are called counter-rotating terms. These terms do not conserve  the total number of excitations in the system: the terms determined by  $a\sigma_-^j$ describe the process in which a photon is annihilated in the cavity mode as the atom makes a downward transition, while the terms  $a^{\dagger}\sigma_+^j$ signify that a photon is created in the cavity mode as the atom makes an upward transition. If the counter rotating terms, $a^{\dagger}\sigma_+^j$ and $a\sigma_-^j$ for $j=A,B$, are ignored (i.e., when the RWA is made), the Hamiltonian takes the form
\begin{eqnarray}\label{rwaHamilton}
H_{RWA}&=&\frac{1}{2}\omega_0(\sigma_z^A+\sigma_z^B)+\omega a^{\dagger}a+g(\sigma_+^Aa+\sigma_-^Aa^{\dagger})\nonumber\\
& &+g(\sigma_+^Ba+\sigma_-^Ba^{\dagger}).
\end{eqnarray}
The RWA is valid for small detuning, i.e., $|\Delta|=|\omega_0-\omega|<<\omega_0,\omega$ and for weak couplings, $g<<\omega_0,\omega$.

The master equation~(\ref{mastereqn}) has no analytical solution  neither for RWA nor for non-RWA Hamiltonians~\cite{fjl,jlf}, but can be solved numerically for different initial states. In the present work, we will restrict ourselves to the pure product state with zero initial excitation, i.e., $\rho_S(0)=\left|g_A,g_B,0\right\rangle\left\langle g_A,g_B,0\right|$. Here $\left|g_A,g_B\right\rangle$ denotes that the atoms, $A$ and $B$, are in their ground states and $\left|0\right\rangle$ signifies that there is no photon inside the cavity.
\section{Correlation Measures: Entanglement and Quantum Discord}
The reduced density matrix of the atoms $A$ and $B$ can be calculated by tracing the density operator, $\rho_S$, over the cavity degrees of freedom. For the initial state considered in the present work, the reduced density matrix of the atoms in the two-qubit standard basis $\left|1\right\rangle\equiv\left|e_A,e_B\right\rangle,\left|2\right\rangle\equiv\left|e_A,g_B\right\rangle,\left|3\right\rangle\equiv\left|g_A,e_B\right\rangle$ and $\left|4\right\rangle\equiv\left|g_A,g_B\right\rangle$ has the X structure:
\begin{eqnarray}
\label{xmatrix}
\rho_{AB}=\left (\begin{array}{cccc} \rho_{11}  & 0 & 0  & \rho_{14} \\ 0  & \rho_{22} & \rho_{23}  & 0 \\ 0  & \rho_{32} & \rho_{33}  & 0 \\ \rho_{41}  & 0 & 0  & \rho_{44} \end{array} \right) \ .
\end{eqnarray}
Based on symmetry considerations, it can be shown that the dynamics in Eq.~(\ref{mastereqn}) preserves the X-form of the density matrix~\cite{rau}. By using the density matrix~(\ref{xmatrix}), the quantum correlations between the atoms as measured by entanglement and quantum discord can be calculated. We adopt Wootters' concurrence~\cite{wooters} as  entanglement measure which is a normalized measure of entanglement and gives 0 for separable states and 1 for maximally entangled (Bell) states. For the density matrix~(\ref{xmatrix}), the concurrence function reads 
\begin{eqnarray}\label{concurrence}
C(t)=2\max\{0,C_1(t),C_2(t)\}, \nonumber\\
C_1(t)=|\rho_{23}|-\sqrt{\rho_{11}\rho_{44}}, \quad C_2(t)=|\rho_{14}|-\sqrt{\rho_{22}\rho_{33}}.\nonumber\\
\end{eqnarray}

On the other hand, quantum discord captures non-classical correlations between two two-level systems that are more general than entanglement~\cite{howz}. It is defined as
\begin{eqnarray}\label{qd}
D(t)=I(t)-J(t),
\end{eqnarray}
where $I(t)=S(\rho_{A})+S(\rho_B)-S(\rho_{AB})$ is the total correlations between the atoms; $S(\rho)=-Tr(\rho\log \rho)$ is the von Neumann entropy and $\rho_A~(\rho_B)$ is the reduced density matrix obtained by tracing $\rho_{AB}$ over the subsystem $B~(A)$. The other quantity $J(t)$ is the measure of classical correlations between the atoms defined as the maximum information one can get about the atom $A$ (or $B$) by performing a set of von Neumann type measurements on the atom $B$~(or $A$)~\cite{fwbac}. Obviously, $J(t)$ and also QD are not symmetric quantities; i.e., they depend on which the measurement is performed. Here we shall consider a set of positive-operator-valued measurements performed on the subsystem $B$~\cite{wlnl}. Recently, analytical expressions for QD of X-state density matrix have been reported~\cite{wlnl,maarga,czyyo}. Here, we will use the results given in Ref.~\cite{wlnl} which are, in fact, equivalent to that of Ali {\it et al.} given in Ref.~\cite{maarga}. According to the results in Ref.~\cite{wlnl}, quantum discord is given as
\begin{eqnarray}\label{qd2}
D(t)=\min\{Q_1,Q_2\},
\end{eqnarray}
where $Q_j=M(\rho_{11}+\rho_{33})+\sum_{i=1}^4\lambda_{AB}^i\log_2\lambda_{AB}^i+P_j$, with $\lambda_{AB}^i$ being the eigenvalues of $\rho_{AB}$, $P_1=M(\tau)$, $P_2=-\sum_{i=1}^4\rho_{ii}\log_2\rho_{ii}-M(\rho_{11}+\rho_{33})$, $\tau=(1+\sqrt{(1-2(\rho_{33}+\rho_{44}))^2+4(|\rho_{23}|+|\rho_{14}|)^2})/2$ and $M(\alpha)=-\alpha\log_2\alpha-(1-\alpha)\log_2(1-\alpha)$ is the binary Shannon entropy function.

For pure states, entanglement of formation and quantum discord are found to be equivalent, while for mixed states such an identification is much more difficult to make. These two measures can disagree on the quantum correlations of a mixed state, for example, it was shown that some unentangled mixed states can carry non-zero quantum discord~\cite{mps}. One should note that although QD measures non-classical correlations that are more general than entanglement, it is not a faithful non-classical correlation measure since it does not vanish only for a state which is strictly classical correlated (see Ref.~\cite{nonclasme}).
\section{Results}\label{results}
In the following, we will investigate the creation of entanglement and quantum discord between two atoms interacting with a lossy single mode cavity field and initially in a state with zero excitation. To do this, we will solve the master equation~(\ref{mastereqn}) for the non-RWA Hamiltonian and we will use Eqs.~(\ref{concurrence}) and~(\ref{qd2}) to calculate concurrence and QD, respectively.  In our calculations, we will fix the detuning $\Delta=\omega_0-\omega=0.01\omega$.

A brief outline of the numerical procedure to solve Eq.~(\ref{mastereqn}) should be given. Due to excitation number being a conserved quantity in Jaynes-Cummings model, the Hamiltonian is block-diagonal in the Hilbert space of $C^2\otimes C^2\otimes F^\infty$ (where $C^2$ and $F^{\infty}$ indicate the qubit and $\infty$-dimensional Fock space for the field, respectively), and diagonalizing blocks corresponding to total excitations of the system gives the analytic solution of the problem. In the full-model, excitation conservation is no longer valid and one should diagonalize an infinite dimensional Hamiltonian. In practice, for the considered initial state we have done a convergence study of the Fock space dimension of the  cavity field~\cite{arm,fjl,jlf,seke}. In this work, we have considered basis vectors of type  $\left|i_A,j_B,n\right\rangle$ where $i,j=e,g$ and $n=0,1,2,...,N$. The converge criterion was considered as the absolute value of the system density matrix elements was smaller than $10^{-10}$ which was found to be satisfied approximately for $N=38$ for the largest atom-field coupling constant and the smallest cavity decay considered in the present work; for the results reported in the remainder of the text, we have taken into account all the basis vectors  $\left|i_A,j_B,n\right\rangle$ where $i,j=e,g$ and $n=0,1,2,...,50$. Also, for all the considered cases the basic properties of the atom-atom density matrix, such as positivity, hermiticity and trace preservation, have been checked during the computational process.
\begin{figure}[!hbt]\centering
{\scalebox{0.5}{\includegraphics{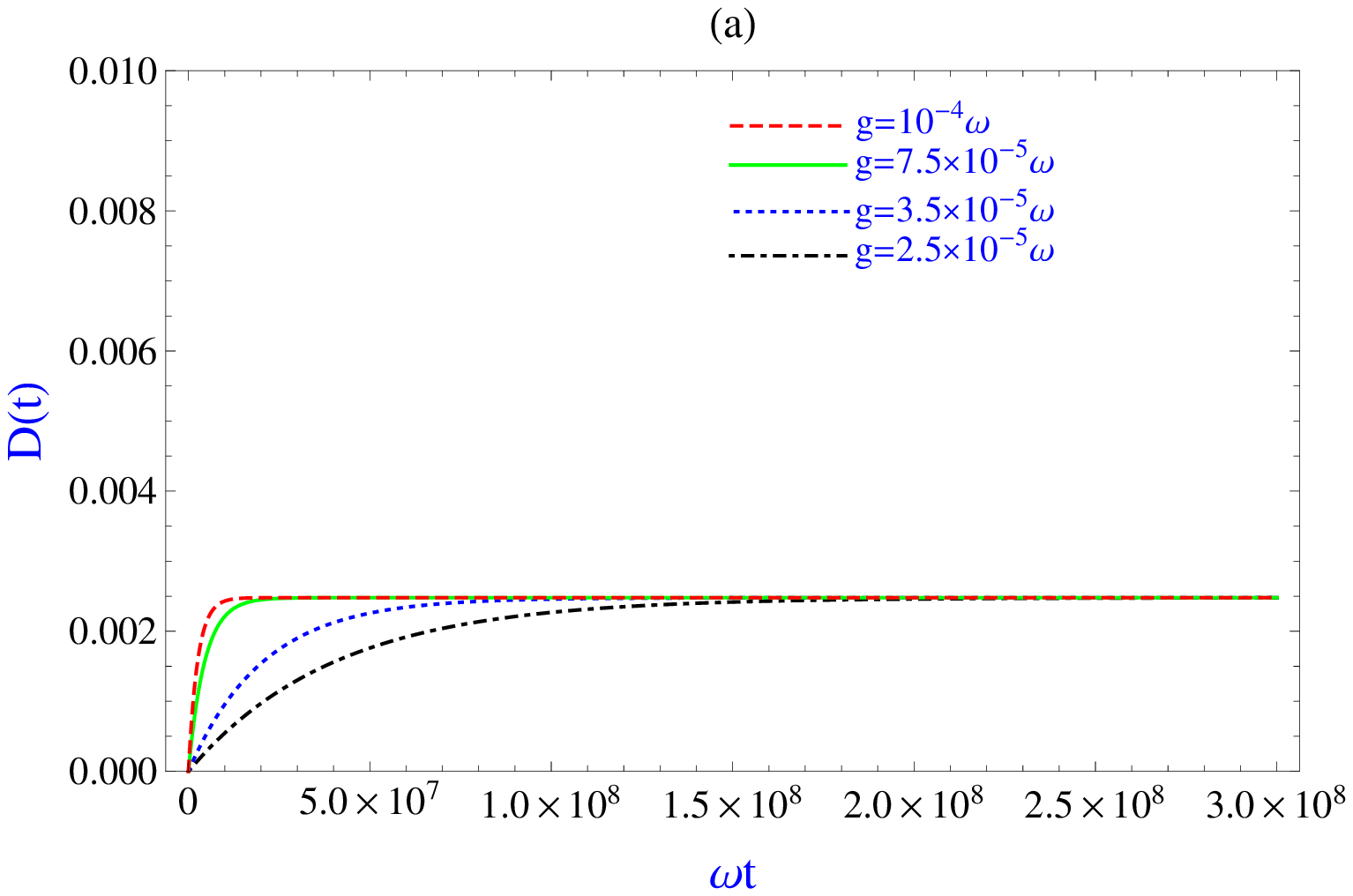}}}
{\scalebox{0.5}{\includegraphics{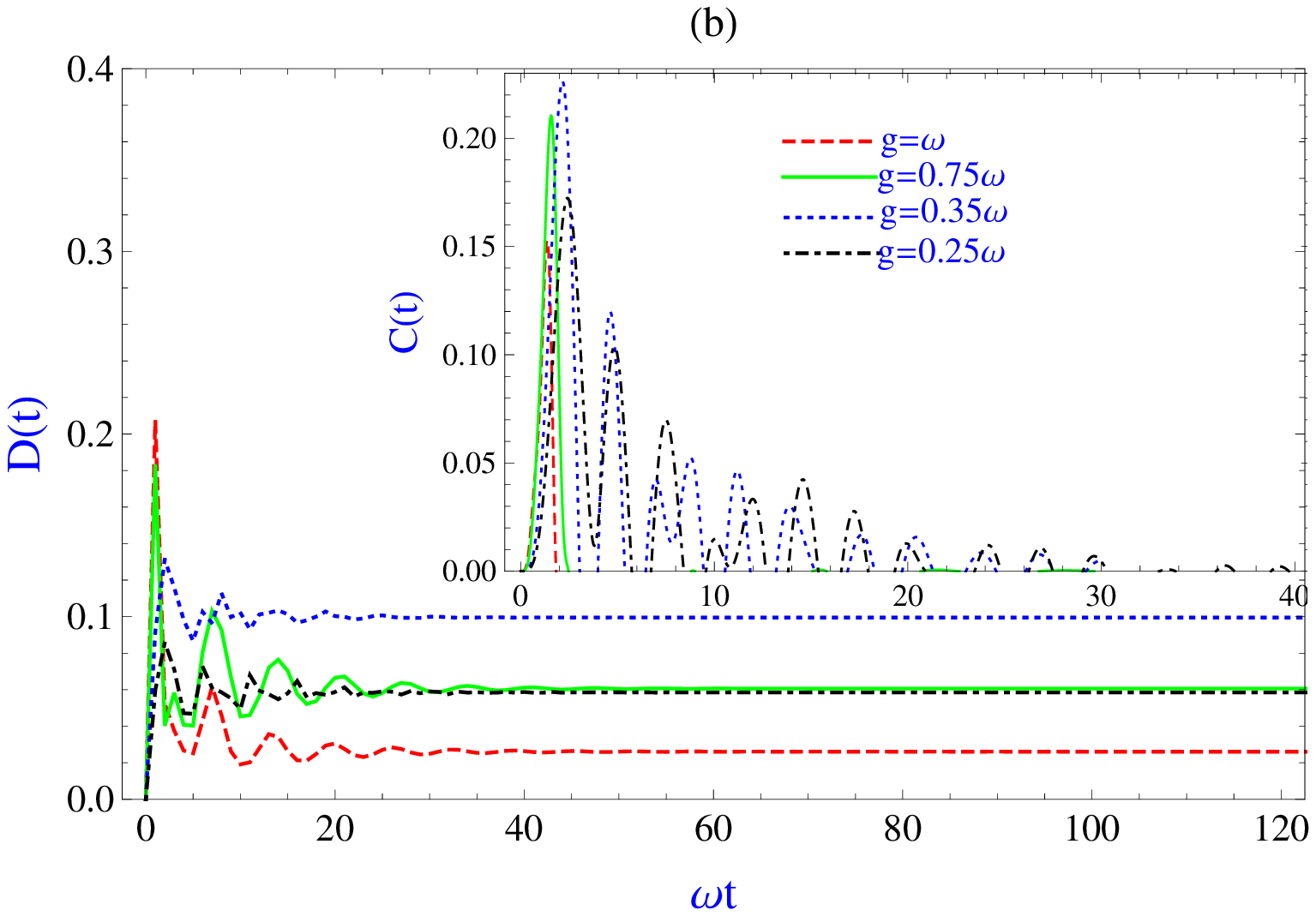}}}
\caption{Effect of atom-field coupling strength, $g$, on the dynamics of quantum discord and concurrence~(inset in~(b)) versus $\omega t$ for $\rho_S(0)=\left|g_A,g_B,0\right\rangle\left\langle g_A,g_B,0\right|$ initial state, $\omega_0=1.01\omega$, $\kappa=0.2\omega$ and for two-qubit Rabi model. Here (a) describes the weak coupling regime and includes plots for $g=1.0\times 10^{-4}\omega$~(red, dashed), $g=7.5\times 10^{-5}\omega$~(green, solid), $g=3.5\times 10^{-5}\omega$~(blue, dotted) and $g=2.5\times 10^{-5}\omega$~(black, dot-dashed), while  (b) indicates strong coupling regime and includes plots for $g=\omega$~(red, dashed), $g=0.75\omega$~(green, solid), $g=0.35\omega$~(blue, dotted) and $g=0.25\omega$~(black, dot-dashed). Note that for weak coupling no entanglement is induced, so they are not plotted here.}
\end{figure}
\begin{figure}[!hbt]\centering
{\scalebox{0.5}{\includegraphics{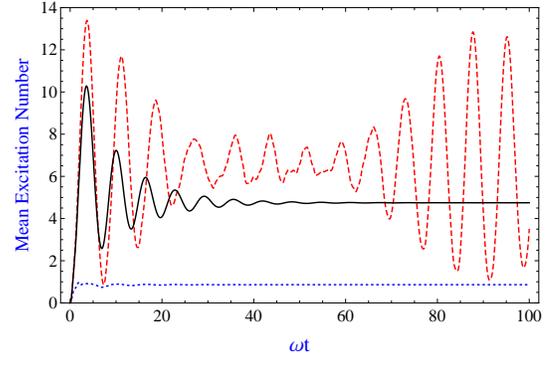}}}
\caption{Time evolution of the mean excitation number in the whole system, $\left\langle N_T\right\rangle_t$~(red-dashed and black-solid), and in the atomic system, $\left\langle N_A\right\rangle_t$~(blue-dotted), versus $\omega t$ for $\rho_S(0)=\left|g_A,g_B,0\right\rangle\left\langle g_A,g_B,0\right|$ initial state, $\omega_0=1.01\omega$, $g=\omega$ and $\kappa=0$~(red-dashed) and $\kappa=0.2\omega$~(black-solid and blue-dotted). Here, $\left\langle N_T\right\rangle_t=\left\langle a^{\dagger}a+\sigma_+^{A}\sigma_-^{A}+\sigma_+^{B}\sigma_-^{B}\right\rangle_t$ and $\left\langle N_A\right\rangle_t=\left\langle \sigma_+^{A}\sigma_-^{A}+\sigma_+^{B}\sigma_-^{B}\right\rangle_t$ where  $\left\langle \ldots\right\rangle_t=Tr\left(\rho_S\{\ldots\}\right)$.}
\end{figure}

First, we investigate the effect of atom-field coupling strength on the time evolution of concurrence and quantum discord. Fig.~1 displays the time-dependent QD and concurrence for the initial state $\rho_S(0)=\left|g_A,g_B,0\right\rangle\left\langle g_A,g_B,0\right|$ with $\Delta=(\omega_0-\omega)=0.01\omega$~ and $\kappa=0.2\omega$. Fig.~1(a) is for the so called weak-coupling regime where $g=1.0\times 10^{-4}\omega$, $g=7.5\times 10^{-5}\omega$, $g=3.5\times 10^{-5}\omega$ and $g=2.5\times 10^{-5}\omega$, while Fig.~1(b) is for the strong coupling regime with $g=\omega$, $g=0.75\omega$, $g=0.35\omega$ and $g=0.25\omega$. It is expected that the difference between RWA and non-RWA evolutions will be more pronounced for the strong coupling parameters. Since RWA dynamics conserve the total excitation number and the considered initial state has zero excitation to start with, both QD and concurrence remain zero at all times. On the other hand, counter-rotating terms~($a^{\dagger}\sigma_+^j$ and $a\sigma_-^j$~for $j=A,B$) will produce virtual excitations which create quantum correlations between the atoms as can be seen from Figs.~1(a) and~1(b). Even in the weak coupling regime, one can see the effect of counter-rotating terms as displayed in Fig.~1(a); non-RWA dynamics lead to steady-state non-zero quantum discord~($D=0.0025$), albeit very small. The magnitude of $g$ in this regime seems to control the speed of approaching the steady-state QD value rather than the magnitude of the asymptotic QD. The dynamics of QD and concurrence in the strong-coupling regime under non-RWA dynamics is richer than the one under the weak coupling; both concurrence and QD are found to be non-zero as displayed in Fig.~1(b). The induced entanglement goes through a series of sudden deaths and births and finally dies out with an overall lifetime inversely proportional to the coupling constant $g$. In contrast, quantum discord always approaches a non-zero asymptotic value which has no monotonic relation with $g$. The observed behavior of QD suggests that not all quantum correlations created by virtual excitations are lost in the dissipative dynamics of the atom-cavity system. Note that QD approaches its steady value much faster for strong coupling than for weak coupling. In Fig.~2, we display the average excitation number $\left\langle N_T\right\rangle_t=\left\langle a^{\dagger}a+\sigma_+^{A}\sigma_-^{A}+\sigma_+^{B}\sigma_-^{B}\right\rangle_t$ in the system as a function of dimensionless time, $\omega t$, at $g=\omega$ for $\kappa=0$ and $\kappa=0.2\omega$, along with average excitation number, $\left\langle N_A\right\rangle_t=\left\langle \sigma_+^{A}\sigma_-^{A}+\sigma_+^{B}\sigma_-^{B}\right\rangle_t$, in the atomic subsystem for $g=\omega$ and $\kappa=0.2\omega$. Due to the collective radiation inhibition effects, the so called "virtual photons" produced by the counter-rotating terms  remain in the cavity despite the strong dissipation~\cite{seke}~(see solid line in Fig.~2). Fig.~2 also shows that  the mean excitation number of the atomic subsystem for leaky cavity is also frozen in the long time limit which is nearly 1. This leads to the appearance of steady states which can have high QD. Moreover, as can be seen from this figure, $\left\langle N_T\right\rangle_t$ is non-zero for both $\kappa=0.2\omega$ and the non-dissipative cavity~($\kappa=0$). Non-zero $\left\langle N_T\right\rangle_t$ is sometimes claimed to be due to the dissipation inhibiting the destruction of virtually created photons in the cavity~\cite{wddv,dodonov2} which seems to be not the case because even for $\kappa=0$, $\left\langle N_T\right\rangle_t$ has a high value.
\begin{figure}[!hbt]\centering
{\scalebox{0.5}{\includegraphics{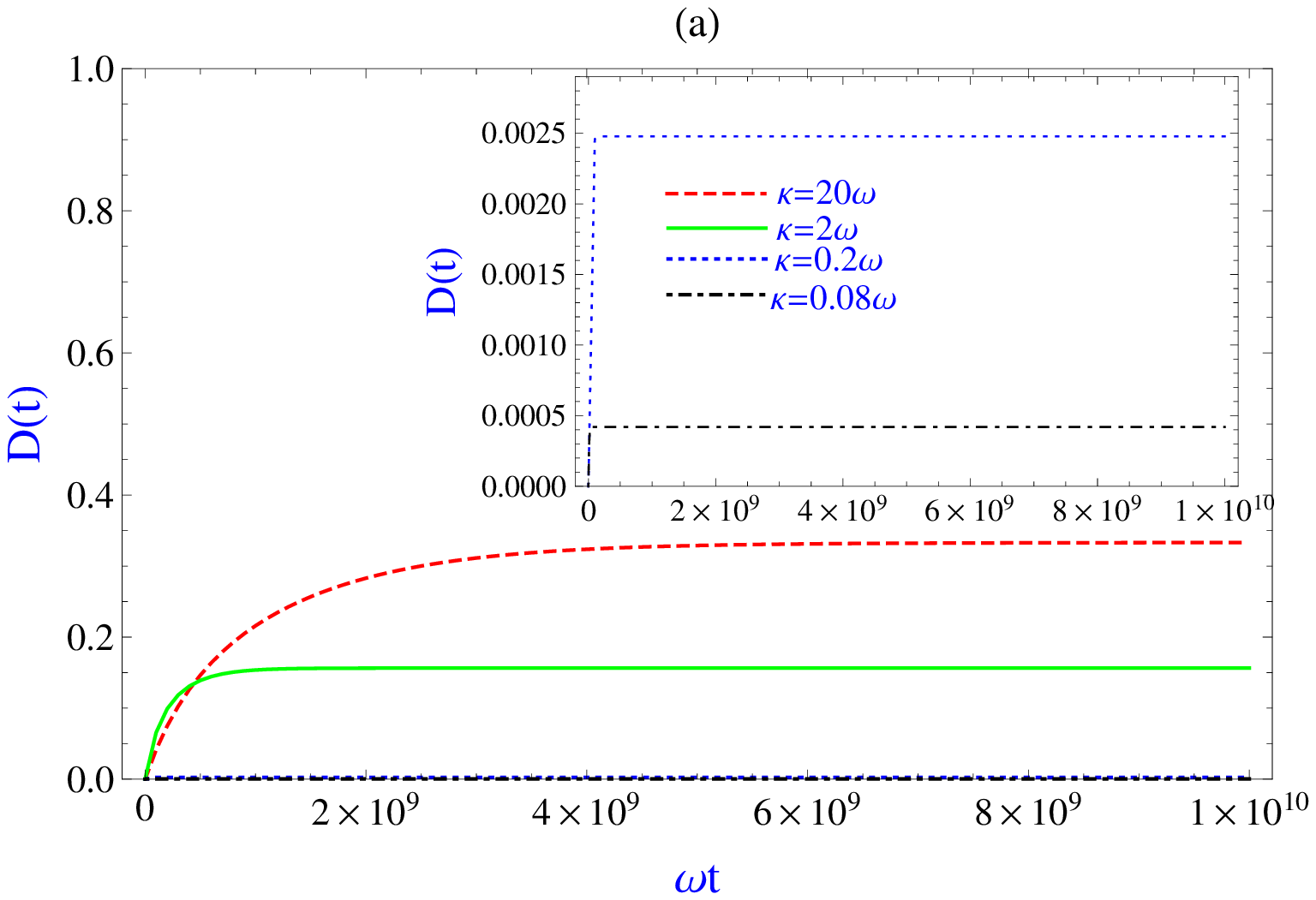}}}
{\scalebox{0.5}{\includegraphics{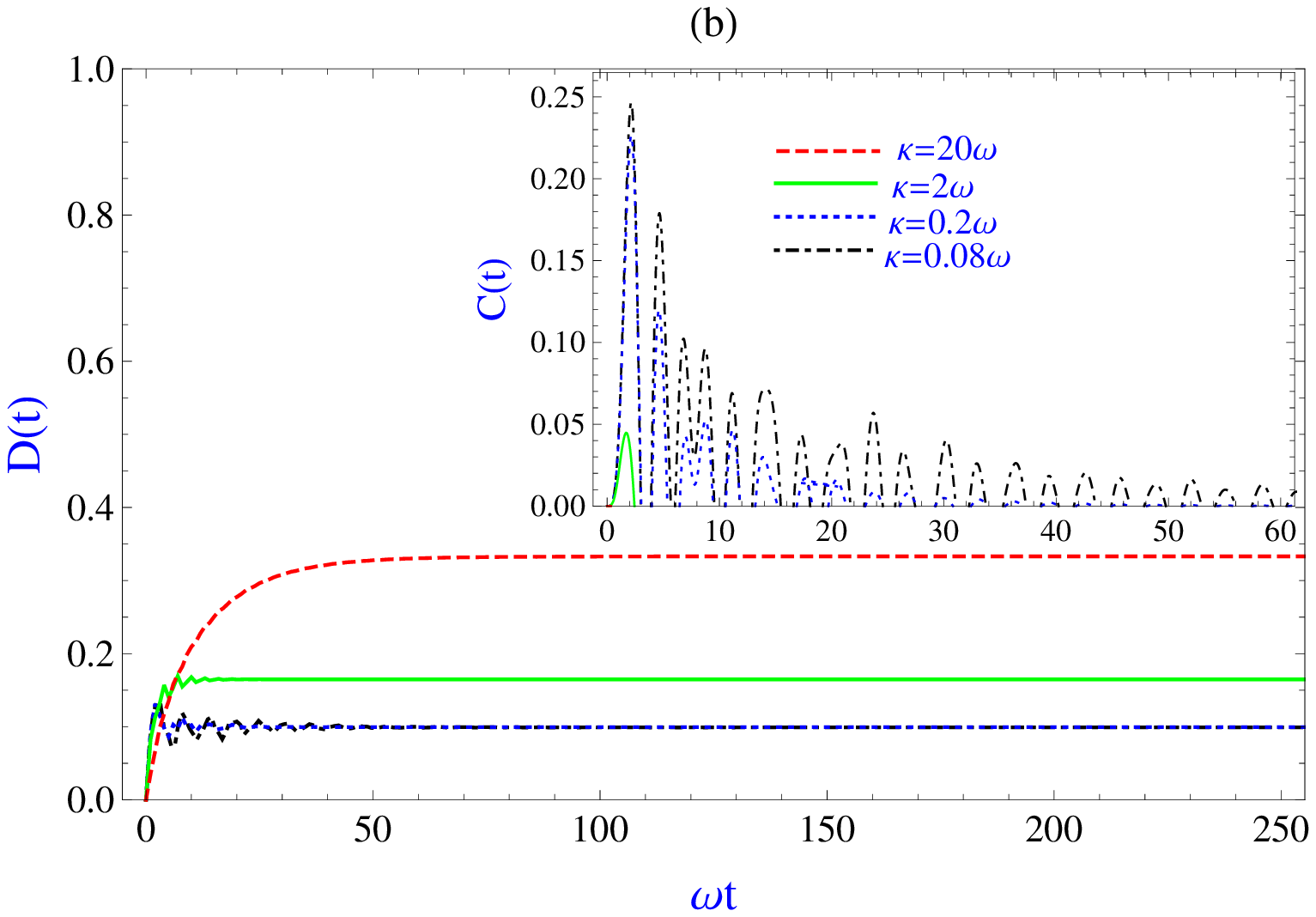}}}
\caption{Effect of cavity decay on the dynamics of quantum discord and concurrence~(inset in~(b)) versus $\omega t$ for $\rho_S(0)=\left|g_A,g_B,0\right\rangle\left\langle g_A,g_B,0\right|$ initial state, $\omega_0=1.01\omega$, $\kappa=0.08\omega$~(black,  dot-dashed), $\kappa=0.2\omega$~(blue, dotted), $\kappa=2\omega$~(green, solid), $\kappa=20\omega$~(red, dashed)  and without the RWA. Here (a) describes the weak coupling regime and includes plots for $g=3.5\times 10^{-5}\omega$, while  (b) indicates strong coupling regime and includes plots for  $g=0.35\omega$. Note that for $g=3.5\times 10^{-5}\omega$ and all considered cavity decay rates as well as for $g=0.35\omega$ and $\kappa=20\omega$, no entanglement is induced, thus they are not plotted here. 	The inset in (a) shows QD versus $\omega t$ for small QD region.}
\end{figure}

Now, we consider the effect of cavity decay on the dynamics of concurrence and QD for the same initial state~($\left|g_A,g_B,0\right\rangle$) and detuning~($\Delta=0.01\omega$) in weak~( $g=3.5\times 10^{-5}\omega$) and strong~($g=0.35\omega$) coupling regimes. We display the QD and concurrence at $\kappa=0.08\omega$, $\kappa=0.2\omega$, $\kappa=2\omega$ and $\kappa=20\omega$ for weak and strong coupling regimes in Figs.~3(a) and~3(b), respectively. Increasing the cavity decay rate is found to decrease the maximum of induced entanglement and to hasten its death at strong coupling regime, as expected~(see the inset in Fig.~3(b)). On the other hand, the decay rate dependence of quantum discord is surprising for both the weak and strong coupling parameters as can be seen from Figs.~3(a) and~3(b); the increasing the cavity decay rate is found to enhance the steady state QD. For high values of $\kappa$~($\kappa=2\omega$ and $\kappa=20\omega$) QD approaches relatively high steady-state value independently of whether $g$ is large or small, while for lesser decay rates~($\kappa=0.08\omega$ and $\kappa=0.2\omega$) the steady-state QD is quite different for the weak and strong coupling cases; the steady state QD is prominently high for $g=0.35\omega$ compared to the one for $g=3.5\times10^{-5}\omega$. One should note that the steady state is reached in much longer times for the weak-coupling case compared to that of the strong-coupling. These findings can be explained based on the results in Ref.~\cite{seke} where the author studied the effects of counter rotating terms on the expectation value of the population inversion operator~(energy shift) for N identical two level atoms interacting with a single mode cavity field in a leaky cavity~(dissipative Dicke Model). It was found  that in very bad-quality cavities, significant steady-state energy shifts can be obtained which increases with the increase in $\kappa$, while these shifts do not depend on the size of the atom-field coupling constant. On the other hand, it was also demonstrated that these energy shifts are small and mainly depend on $g$ for high-quality cavities. From the above results, the large cavity damping is found to increase the contribution of counter-rotating terms and provides us a way to see the effects of virtual processes at low atom-field coupling constants. The role of $\kappa$ in the creation of long time non-zero QD can be better understood by looking at the atom-atom density matrix; for example, for $\Delta=0.01\omega$, $g=0.35\omega$ and $\kappa=0.08\omega$, the steady state density matrix is equal to
\begin{eqnarray}
\rho_{AB}(\kappa=0.08\omega)=\left (\begin{array}{cccc} 0.014  & 0 & 0  & 0.074 \\ 0  & 0.073 & 0.073  & 0 \\ 0  & 0.073 & 0.073  & 0 \\ 0.074  & 0 & 0  & 0.84 \end{array} \right) \nonumber,
\end{eqnarray}
with steady discord $D=0.1$, while for $\kappa=20\omega$,
\begin{eqnarray}
\rho_{AB}(\kappa=20\omega)=\left (\begin{array}{cccc} 0.32  & 0 & 0  & 3.1\times 10^{-5} \\ 0  & 0.17 & 0.17  & 0 \\ 0  & 0.17 & 0.17  & 0 \\3.1\times 10^{-5}  & 0 & 0  & 0.34 \end{array} \right)\nonumber,
\end{eqnarray}
with steady discord $D=0.33$. Comparing these two density matrices one can note that the populations $\rho_{11}=\left|e_A,e_B\right\rangle\left\langle e_A,e_B\right|$, $\rho_{22}=\left|e_A,g_B\right\rangle\left\langle e_A,g_B\right|$ and $\rho_{33}=\left|g_A,e_B\right\rangle\left\langle g_A,e_B\right|$ and the coherence $\rho_{23}=\left|e_A,g_B\right\rangle\left\langle g_A,e_B\right|$ increase with the increase of $\kappa$, while the population $\rho_{44}=\left|g_A,g_B\right\rangle\left\langle g_A,g_B\right|$ and the coherence $\rho_{14}=\left|e_A,e_B\right\rangle\left\langle g_A,g_B\right|$ decrease. Such a change can enhance the steady state quantum discord. One can conclude that the increase in the cavity decay can decrease the chance of atoms to interact with cavity field and the atomic system is frozen in a state with high quantum correlations before it has a chance to decay to the ground state~\cite{hxyz}. It is interesting to note that in the very bad quality cavity case~($\kappa/\omega>20$), the steady atomic density matrix is saturated and can be nearly written as an equal weighted sum, $\rho_{AB}(\kappa/\omega>20)=0.333\left(\left|e_A,e_B\right\rangle\left\langle e_A,e_B\right|+\left|g_A,g_B\right\rangle\left\langle g_A,g_B\right|\right)+0.334\left|\Psi^+\right\rangle\left\langle \Psi^+\right|$, where $\left|\Psi^+\right\rangle=\frac{1}{\sqrt{2}}\left(\left|e_A,g_B\right\rangle+\left|g_A,e_B\right\rangle\right)$ is the Bell state, and the $g$-dependence of this atomic steady state becomes negligible as shown in Fig.~3. Although this atomic state contains no entanglement, it has high QD which is nearly $D=0.333$. Recently, it was shown that maximal QD reachable by two separable qubits is 1/3~\cite{mrqd}. As a consequence, it seems that the interplay between losses and virtual processes does not only lead to high steady QD, but also maximizes it for separable states. A similar result has been also obtained in different systems~(see, for example Ref.~\cite{wlnl}). One should note that a negligibly small amount of steady entanglement~($C\approx 2\times 10^{-3}$) exists in the state only for $\kappa=0.08\omega$ that can be noted from the above reduced density matrix.

\section{Conclusion}
We have investigated the effect of counter-rotating interaction terms on the dynamics of entanglement and quantum discord between two qubits interacting with a single mode cavity field in a leaky cavity for a system with zero initial excitation. We have shown that virtual processes can lead to long-lived constant quantum discord even at weak atom-field couplings~($g/\omega\approx 10^{-5}$) and can create non-zero entanglement at strong coupling regime which goes through sudden death and birth processes before permanently vanishing~(for $\kappa>0.08\omega$).

A counter-intuitive finding of the present work is the cavity decay rate dependence of entanglement and quantum discord; while an increase in decay rate leads to a shorter lifetime for entanglement, it increases the steady state value of quantum discord for both strong and weak couplings. Moreover, for high quality cavities~($\kappa=0.08\omega$ and $\kappa=0.2\omega$), the steady state QD is found to be highly interaction strength dependent, while for low-quality cavities~($\kappa=2\omega$ and $\kappa=20\omega$), the dependence of steady state QD on the interaction strength is found to be negligible. In fact, the competition between counter-rotating terms and cavity decay is found to give not only a relatively high value of steady QD, but indeed maximizes it for separable states in the case of very bad quality cavity~($\kappa/\omega>20$).

It is worth mentioning here that the master equation~(\ref{mastereqn}) has been used by a large number of groups in recent years in order to study the cavity decay in Rabi model~\cite{wddv,fjl,jlf,seke}, but its validity has not been proved yet~\cite{wddv}. Nevertheless, if the master equation is indeed applicable, the results reported in this Letter might be relevant for the strong coupling experimental work which is made possible with the recent cavity QED circuit proposals~\cite{aljhe,holstein,scqnmr,etcref}.

\section*{Acknowledgments}

We would like to thank anonymous Referees for constructive remarks.

\end{document}